\documentclass[aps,prl,twocolumn,superscriptaddress]{revtex4}
\usepackage{mathtools}
\usepackage{graphicx}
\usepackage{hyperref}

\newcommand{\abs}[1]{| #1 |}

\usepackage{xcolor}

\begin{document}
\title{Spectral Compression of Narrowband Single Photons with a Resonant Cavity}
\author{Mathias~A.~Seidler}
\affiliation{Centre for Quantum Technologies, National University of Singapore, 3 Science Drive 2, Singapore 117543}

\author{Xi~Jie~Yeo}
\affiliation{Department of Physics, National University of Singapore, 2 Science Drive 3, Singapore 117551}

\author{Alessandro~Cer\`{e}}
\affiliation{Centre for Quantum Technologies, National University of Singapore, 3 Science Drive 2, Singapore 117543}

\author{Christian~Kurtsiefer}
\affiliation{Centre for Quantum Technologies, National University of Singapore, 3 Science Drive 2, Singapore 117543}
\affiliation{Department of Physics, National University of Singapore, 2 Science Drive 3, Singapore 117551}

\email[]{christian.kurtsiefer@gmail.com}

\date{\today}
\begin{abstract}
We experimentally demonstrate a spectral compression scheme for heralded
single photons with narrow spectral bandwidth around 795\,nm,
generated through four-wave mixing in a cloud of cold
$^{87}$Rb atoms.
The scheme is based on an asymmetric cavity as a dispersion medium and a
simple binary phase modulator, and can be, in principle,  without any optical
losses. We observe a compression from 20.6\,MHz to less than 8\,MHz, almost
matching the corresponding atomic transition.
\end{abstract}


\maketitle
\textit{Introduction --}
Efficient atom-light interactions at the single quantum level is at 
the core of several proposals for storing, processing, and relaying quantum
information~\cite{Cirac1997,Briegel1998,Waks2009, Kimble2008}. 
Many of these schemes
require single ``flying'' photons to match the spectrum of atomic transitions~\cite{Wilk2007,Hammerer2010,Lukin2003,Steiner:17}.
Single photons can be emitted from trapped ions~\cite{Keller2004,Almendros2009}, atoms~\cite{Kuhn2002,Srivathsan:2013,park2019polarization} or solid-state systems~\cite{Kurtsiefer2000,Michler2000,Moreau2001}.
However, the spectral width of the generated photons may not always match the
spectral width of the receving systems. Therefore, methods to engineer the
photon spectrum may be required.

The simplest method for this is to passively filter the spectrum of
bright broadband sources \cite{Meyer-Scott2017, Schuck2010},
with a sometimes significant reduction of brightness, makeing photon-atom
interaction experiments that require a high interaction
rate~\cite{Duan2001b,Blinov2004} difficult.
More advanced methods to manipulate the spectrum of single photon
sources to match that of 
atomic transitions include restricting the spectral mode of emitters with cavities~\cite{Moreau2001,Kuhn2002,McKeever2004,Wolfgramm:08}, or 
using electromagnetically induced transparency in atomic
ensembles~\cite{Eisaman2005,zhu2017bright} in the source mechanism altogether.
Another reported approach changes the photon spectrum with a gradient echo quantum memory~\cite{Buchler2010}.
As spectral filtering or engineering of the photon generation mechanism may
not always be possible, it would be desirable to modify the spectrum of a given
photon source while maintaining the brightness.

Here, we demonstrate such a technique that compresses the spectral bandwidth
of already narrowband single photons, which in principle, maintains the photon rates. 
The technique is based on the ideas of time-lenses invented for temporal
imaging~\cite{Kolner1989,Kolner1994}, where the temporal and spectral
characteristics of ultrafast electromagnetic
pulses~\cite{Lavoie2013,Karpinski2017,Li2017} are manipulated. It turns out
that single photon states can be manipulated in a similar way, complementing
the techniques for lossless temporal envelope manipulation of narrowband
single photon states demonstrated
in~\cite{Srivathsan2014,PhysRevLett.123.133602}.

Spectral compression of single photon wave packets is achieved in two
steps:  First, the wave
packet is spread out in time such that the width of its envelope is compatible
with a narrow spectrum; this can be done using a dispersive element that
spreads out different frequency components of the wave packet in time,
effectively generating a chirped wave packet. 
In the second step, a time-dependent phase shift is applied. 
This step changes the spectral energy distribution of the wave packet. 

Previous time-lens schemes used to spectrally compress ultrashort pulses used
optical fibers or diffraction gratings as dispersive
media~\cite{Lavoie2013,Karpinski2017,Li2017}. For photonic wave packets
interacting with single emitters like atoms or molecules, the relevant
spectral bandwidth is on the order of a few MHz, which makes the use of fibers
or gratings as dispersive elements impractical due to losses in
the required very long fibers.
We overcome this problem by
using the dispersive properties of an optical cavity instead. While the
dispersion in optical cavities can be much larger,
the process then requires a different time-dependent phase shift in the second
step to complete the spectral compression process.

\begin{figure}
  \includegraphics[width=\columnwidth]{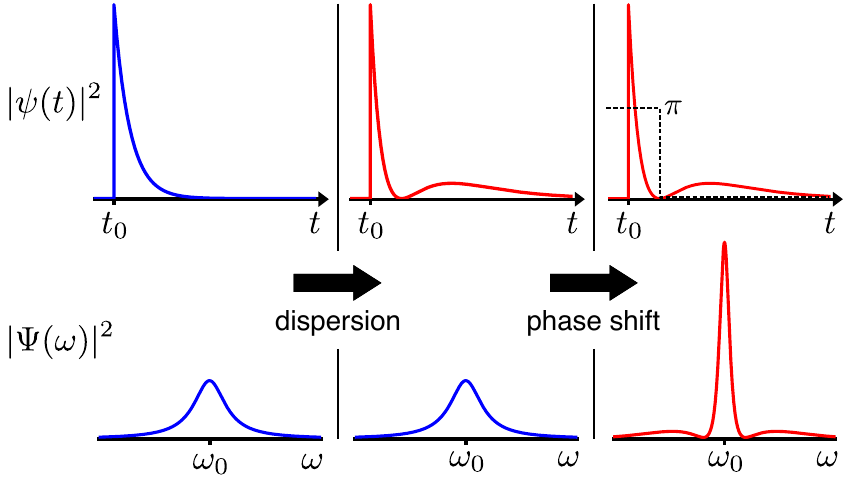}
    \caption{\label{fig:concept}
      Concept of spectral compression. The top row shows temporal intensity
      profiles $\abs{\psi(t)}^{2}$ in various stages of the spectral
      compression, the bottom row the corresponding power spectra $\abs{\Psi(\omega)}^{2}$.
      The initial pulse is dispersed by a cavity, leading to a new temporal
      shape, but an unchanged spectrum. An electro-optical modulator (EOM) manipulates the phase
      $\phi^{\prime}(t)$ of the pulse which leads to a narrower spectrum.}
\end{figure}
\textit{Theory --} To understand the spectral compression scheme, we start with
an initial single photon wave packet, described by an envelope
$\left|\psi(t)\right|^2$ of its intensity in time, and its corresponding  power spectrum
$\left|\Psi(\omega;\, \omega_0, \Gamma_p)\right|^2$,
connected by the Fourier transform~$F$:
$\Psi(\omega)=F\left[\psi(t)\right]$. The nearly-monochromatic wave packet
shall be characterized by a central frequency~$\omega_0$ and a spectral width~$\Gamma_p$.
The spreading out of the wave packet in time is accomplished by reflection off
an asymmetric cavity, with an input/output coupler with a low transmission,
and a second high-reflective mirror, similar to the setup used in~\cite{Srivathsan:2014}.

If the losses in the cavity are negligible compared to the transmission of the
coupling mirror, and the 
cavity linewidth~$\Gamma_c$ and photon bandwidth~$\Gamma_p$ are much smaller
than free spectral range of the cavity,
the action of the cavity to a wave packet near its resonance~$\omega_c$ can be
described by a transfer function
\begin{equation}\label{eq:cavity_tf}
    C(\omega;\, \omega_c, \Gamma_c) \approx
    -\frac{\Gamma_c + i\: 2 (\omega-\omega_c)}
    {\Gamma_c - i\: 2(\omega-\omega_c)}
    \,,
\end{equation}
which modifies the incoming spectral wave packet $\Psi(\omega;\, \omega_0,
\Gamma_p)$ to a new one,
\begin{equation}\label{eq:Psi_dispersed}
    \Psi^{\prime}(\omega;\, \Delta\omega, \Gamma_c, \Gamma_p) =\:
    \Psi\left(\omega;\, \omega_0, \Gamma_p\right)\:C\left(\omega;\, \omega_c,\Gamma_c\right)\,,
\end{equation}
where~$\Delta\omega=\omega_0 - \omega_c$ is the detuning between the wave packet and the cavity resonance.
For a lossless cavity, this wave packet has the same power spectrum
as~$\Psi(\omega)$ because $\left| C(\omega;\, \omega_c, \Gamma_c)\right|^2 = 1$.
The temporal envelope of the reflected wave packet, obtained through the inverse Fourier transform~$F^{-1}$,
\begin{equation}\label{eq:psi_dispersed}
    \psi^{\prime}(t;\,\Delta\omega, \Gamma_c, \Gamma_p)
    =\:
    F^{-1}\left[\Psi^{\prime}(\omega;\,\Delta\omega, \Gamma_c, \Gamma_p)\right]\,,
\end{equation}
is now broader in time, and has acquired a time-dependent phase~
$\phi^{\prime}(t;\,\Delta\omega, \Gamma_p, \Gamma_c)$\,.

Similar to Fourier-transform limited pulses, where the time-bandwidth product is minimized by a frequency-independent spectral phase,
we can reduce the spectral bandwidth
of the heralded single photon by removing any time-dependent phase.
This is done by applying a time-dependent phase shift
\begin{equation}\label{eq:eom_dispersion}
    \phi_e(t)= -\phi^{\prime}(t;\;\Delta\omega, \Gamma_p, \Gamma_c)\,,
\end{equation}
resulting in the spectrally-compressed wave packet
\begin{equation}\label{eq:compressed_psi}
    \psi^{\prime\prime}(t;\;\Delta\omega, \Gamma_p, \Gamma_c) =
    \psi^{\prime}(t;\;\Delta\omega, \Gamma_p, \Gamma_c) \: e^{i \phi_e(t)}\,.
\end{equation}

To quantify the compression, we compare the spectral widths before and after
the compression obtained from the respective power spectra 
obtained through a Fourier transform of Eq.~(\ref{eq:compressed_psi}).

\begin{figure}
 \includegraphics[width=\columnwidth]{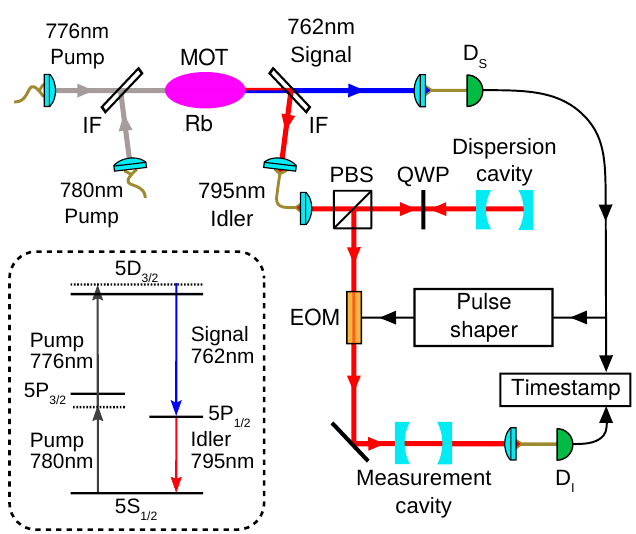}
 \caption{\label{fig:schematic}
   Schematic setup for generation and spectral compression of heralded single photons. 
   D$_{S,I}$: single-photon detectors, EOM: electro-optical modulator, PBS: polarizing beam splitter, QWP: quarter-wave plate, IF: interference filter. Inset: energy level scheme for four-wave mixing in $^{87}$Rb.}
\end{figure}
We now consider the specific case of a heralded single photon emerging from an
atomic cascade decay, where we intend to compress the idler photon (see inset
of Fig.~\ref{fig:schematic}).
Detection of a signal photon projects the field in the idler mode into the heralded state
\begin{equation}\label{eq:heralded_photon}
    \psi(t) = \sqrt{\Gamma_p}\:e^{-\frac{\Gamma_p}{2}(t - t_0) }
    \Theta(t - t_0)
    \,,
\end{equation}
where~$t_0$ and $t$ are the detection times of the signal and idler photons, respectively.
The exponential decay with the constant~$\Gamma_p$ is a characteristic of the spontaneous process,
while
the Heaviside step function~$\Theta$ is a consequence of the well-defined time order of the cascade decay process. For simplicity, we set~$t_0=0$.

This temporal profile corresponds to a Lorentzian power spectrum for the idler
photons, and its bandwidth is described by the full-width half maximum $\Gamma_p$, 
which also corresponds to the spectral window containing 50\% of the total pulse energy.
However, the compressed spectrum
$|F^{-1}\left[\psi^{\prime\prime}(t)\right]|^2$ has multiple maxima, and is
distinctly different from distributions where the full-width half maximum
naturally quantifies the bandwidth.
Hence, we instead define bandwidth as the smallest spectral width containing 50\% of the total pulse energy,
as this definition of bandwidth is compatible for both a Lorentzian and a generic spectrum.

To obtain the optimal cavity parameters, we numerically minimize
the bandwidth of the compressed photon spectrum.
We find that the maximal compression is achieved by a resonant cavity $\Delta\omega=0$ with a bandwidth of $\Gamma_c \approx \Gamma_p/4$.
Under these conditions, the compressed single photon time envelope can be written as 
\begin{equation}\label{eq:comp_psi_exp}
    \psi^{\prime\prime}(t) =e^{-i \phi^{\prime}(t)} \:
    \sqrt{\Gamma_p}\,
    \frac{2\Gamma_c \,e^{-\frac{\Gamma_c}{2} t} -
    (\Gamma_p +\Gamma_c)\, e^{-\frac{\Gamma_p}{2} t}}
    {\Gamma_p -\Gamma_c}\,\Theta(t)\,,
\end{equation}
with a phase function
\begin{equation}\label{eq:phase_function}
    \phi^{\prime}(t) = \pi\,
    \Theta\left(t - 2
    \frac{\log \left(\frac{\Gamma_p +\Gamma_c}
        {2 \Gamma_c}\right)}
    {\Gamma_p -\Gamma_c}
    \right)\,.
\end{equation}
This is a step function changing the phase by $\pi$, with the transition occurring at the minimum of the dispersed photon's temporal intensity profile.
The narrowest bandwidth achievable with compression based on an asymmetric
cavity with this strategy is $\sim\!0.3\Gamma_p$; the temporal envelopes and
power spectra shown in  Fig.~\ref{fig:concept} correspond to this choice.

\textit{Experiment --}
Details of the actual experiment are shown in Fig.~\ref{fig:schematic}.
We generate the time-ordered photon pairs by
four-wave mixing in a cold ensemble of $^{87}$Rb atoms in a cascade level scheme~\cite{Srivathsan:2013}.
Pump beams at 780\,nm and 776\,nm excite atoms from the $5S_{1/2},F=2$ ground level to the $5D_{3/2},F=3$ level via a two-photon transition. The 762\,nm
(signal) and 795\,nm (idler) photon pairs emerge from a cascade decay back to
the ground level, and are coupled to single mode fibers.
Phase matching is ensured with
all four modes propagating collinearly in the same direction.
The two pump have a focus in the cloud with a beam waist of about 400\,$\mu$m. The 780\,nm pump is 55\,MHz blue-detuned from the 5S$_{1/2}$,F=2 to 5P$_{3/2}$,F=3 transition and has an optical power of 0.25\,mW.
The 776\,nm pump
has an optical power of 11.4\,mW, and is tuned such that the two-photon transition to the 5D$_{3/2}$, F=3 state is 5\,MHz blue-detuned.
When the excited atoms decay via the 5D$_{1/2}$, F=2 state back into the
initial ground state, photons with a wavelength of 762\,nm and 795\,nm photon are emitted~\cite{Srivathsan:2013}.

\begin{figure}
  \includegraphics[width=1\columnwidth]{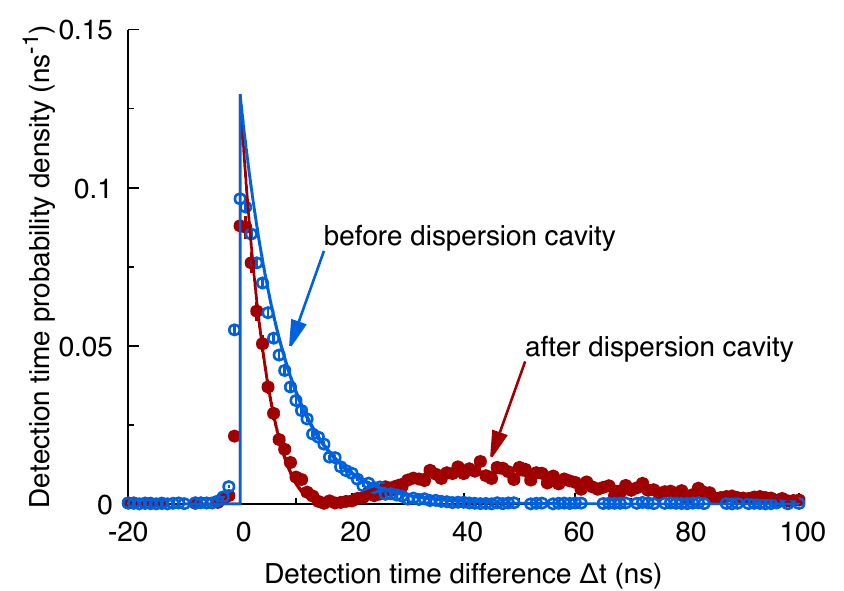}
 \caption{\label{fig:photong_g2s}
   Detection time distribution for the heralded photon
  before (open blue circles) and after
  (filled red dots) the dispersion cavity.
  We fit an exponential decay Eq.~(\ref{eq:heralded_photon}) to the initial time correlation (blue solid line), from which we infer the photon bandwidth $\Gamma_p$.
  The simulated temporal profile after the photon passed through the
  dispersion cavity, (red line, calculated from Eq.~(\ref{eq:Psi_dispersed}),
  matches our experimental data (filled dots) well. }

\end{figure}

After suppressing residual pump light and separating signal and idler photons
into different modes, we collect them into single mode fibers. 
The 762\,nm signal photons are detected with an avalanche photo diode and
herald the presence of 795\,nm idler photons.
The time correlation between the detection in the signal and idler modes
(open 
circles in Fig.~\ref{fig:photong_g2s}) corresponds to the
envelope~$|\psi(t)|^2$ of the intensity in time.

\begin{figure}
 \includegraphics[width=1\columnwidth]{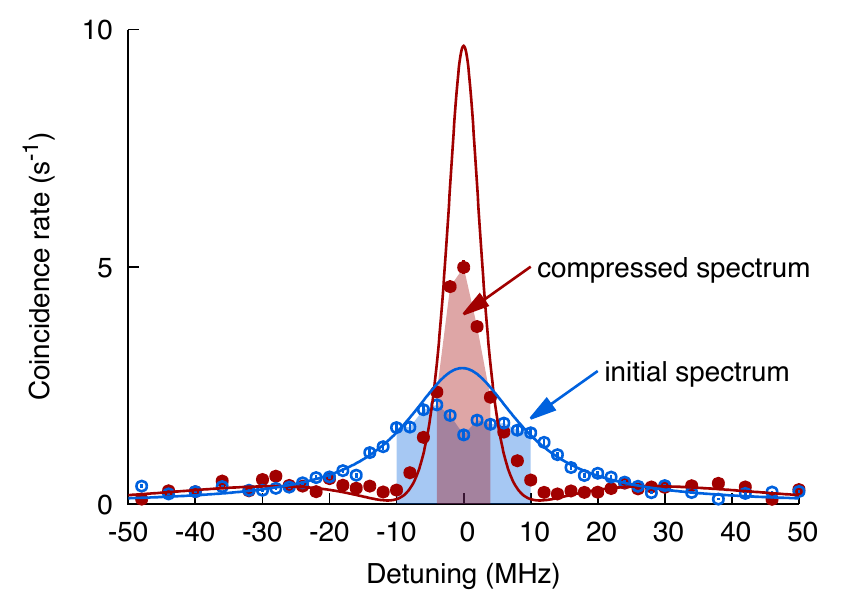}
 \caption{\label{fig:spectrum}
 Spectral profile of heralded photons before (blue) and after (red) spectral compression, obtained by measuring the photon transmission rate through the Fabry-Perot cavity at different cavity detunings. 
The solid lines are calculated from Eq.~(\ref{eq:compressed_psi}), with $\Gamma_p$ inferred from the temporal envelope measurement of the photons from the source, and $\Gamma_c$ by experimentally characterizing the cavity bandwidth.
 Shaded areas cover 50\% of the total power for each spectrum.}
\end{figure}

We measure the initial power spectrum of the wave packet (open circles in Fig.~\ref{fig:spectrum}) 
by correlating it with a the photon rate transmitted through
a Fabry-P\'{e}rot cavity (FP) with linewidth~
$\Gamma_\text{FP}\approx2\pi\times2.6$\,MHz. 
The transmission is recorded at different detunings of the cavity from the
atomic resonance.
The observed spectrum has a full width at half maximum of 20(2)\,MHz, wider
than the atomic line width of 6\,MHz due to collective emission effects in the cloud~\cite{Jen:2012, PhysRevA.98.023835}.

The 795\,nm idler photons are then coupled to the dispersion cavity, with a
coupling mirror of nominal reflectivity~
$R_1=0.97$, and a high reflector with $R_2=0.9995$
separated by 10.1\,cm, corresponding to a free spectral range of
$1.48$\,GHz, and a measured linewidth
$\Gamma_c\approx2\pi\times7.3$\,MHz.
A Pound-Drever-Hall frequency lock keeps the cavity resonant 
to the central frequency of
the photons throughout the experiment.
The measured time envelope of the single photon wave packet after dispersion is
shown as filled 
dots in Fig.~\ref{fig:photong_g2s}.

The spectral compression is completed by applying the temporal phase of Eq.~(\ref{eq:phase_function}), in the form of a phase switch
synchronized to the photon passage through a fiber connected 
electro-optical modulator (EOM).
Since the idler photon is heralded, we use the detection of the signal photon
as a reference signal for triggering the phase switch after an appropriate
time delay.
The phase flip is applied right after the first part of the dispersed photon exits the modulator, and 
the second part starts to propagate through it. This timing is indicated as dashed line in Fig.~\ref{fig:concept}.
We finally measure the compressed photon spectrum by again recording the photon transmission rate through the Fabry-P\'{e}rot cavity, shown as filled dots in Fig.~\ref{fig:spectrum}.


To obtain an initial photon bandwidth $\Gamma_p$, we fit the decaying
exponential term in Eq.~(\ref{eq:heralded_photon}) to the observed coincidence
probability (open circles in Fig.~\ref{fig:photong_g2s}).
The solid red line in Fig.~\ref{fig:photong_g2s} corresponds to an expected temporal
profile of the photon after the dispersion cavity, calculated from
Eq.~(\ref{eq:comp_psi_exp}), with $\Gamma_p$ obtained from the fit of the
initial photon shape, and the cavity linewidth $\Gamma_c\approx2\pi\times7.3$\,MHz measured earlier. 
The observed temporal envelope after the dispersion cavity (full dots in
Fig.~\ref{fig:photong_g2s}) agrees very well with the expected profile.

The measured spectral profiles before and after compression are shown in Fig~\ref{fig:spectrum}.
The spectrum of the uncompressed photons (open circles) exhibits a dip around the central
frequency, which we attribute this to a reabsorption of the generated photons in our source.
We model this by combining a Lorentzian with an absorption model 
\begin{equation}\label{eq:reabsorption}
    S(\omega; OD, \Gamma_p, \Gamma_{a}) = \frac{A}{\pi}\ \frac{2\,\Gamma_p}{4\,\omega^2 + \Gamma_p^2} \ 
    e^{-\text{OD}\frac{\Gamma_a^2}{4\omega^2 + \Gamma_a^2}}\,,
\end{equation}
with the optical density OD of the cloud, and a scaling factor $A$ as free fitting parameters. The value of
$\Gamma_p$ was determined by the previous fit, and $\Gamma_{a}$ is set to 6.06\,MHz, the atomic line width of the $5S_{1/2}\rightarrow5P_{1/2}$ transition.
We fit this model to our data points and plot the 
Lorentzian part (without the self-absorption) as the blue line in Fig.~\ref{fig:spectrum}.
We rescale the 
 expected compressed power spectrum $|\Psi(\omega)|^2$ using the fitted value for the scaling factor $A$, shown as the red line in Fig.~\ref{fig:spectrum}.

\textit{Discussion --}
By definition, spectral compression reduces the width of a spectral
distribution, resulting in an increased photon rate/intensity at the central
frequency.
In our experiment, we find a bandwidth of 20(2)\,MHz for the initial,
and 8(2)\,MHz for the compressed photon.
This almost matches the natural D transition linewidth of 6\,MHz in $^{87}$Rb.
The maximal photon transmission through the spectroscopic cavity is increased
by a factor of 2.39(4), 
indicating a successful spectral compression of narrowband photons.

The compressions mechanism is, in principle, lossless since both cavity and phase modulators can have arbitrarily low losses. 
In our experiment, we incur a total photon loss of about 79\% due to the compression optics. 
We measure an optical transmission of 60\% through the dispersion optics (PBS, QWP, dispersion cavity, fiber coupling), and 35.7\% through the fiber-based EOM. 
The photon loss can be significantly reduced by replacing the fiber-based EOM with a free-space EOM.

The optimal spectral compression of a photon with bandwidth $\Gamma_{p}$ in
the cavity-based scheme is achieved if the dispersion cavity has a bandwidth of $0.25\,\Gamma_p$.
Since the amount of spectral compression is limited by the dispersion
mechanism, dispersion engineering of structured dielectric media~\cite{RevModPhys.78.455,Li_2018,Joannopoulos_2008} or multiple combined optical cavities may allow a further increase of spectral compression.
This method is not limited to the atomic system in our experiment -- it can be
adapted to a wide range of wavelengths and spectral widths, and therefore even
allow to 
match the spectral properties to different types of quantum systems e.g. in a hybrid quantum network~\cite{Kurizki3866}.

We thank Adrian Nugraha Utama for useful discussions on the theoretical
modeling.
This work was supported by the Ministry of Education in Singapore. 

\bibliographystyle{apsrev4-1}
%

\end{document}